Title

**Ejecta emplacement as the possible origin of Ryugu's equatorial ridge**

Authors

**Ren Ikeya [a] and Naoyuki Hirata [a, *]**

* Corresponding Author E-mail address: hirata@tiger.kobe-u.ac.jp

**Authors' affiliation**

[a] Graduate School of Science, Kobe University, Kobe, Japan.

**Proposed Running Head: Ejecta feed Ryugu's equatorial ridge**

Editorial Correspondence to:
Dr. Naoyuki Hirata
Kobe University, Rokkodai 1-1 657-0013
Tel/Fax +81-7-8803-6566

**Keywords**
Asteroid
Impact processes
Geological processes

**Highlights**

- We propose a new mechanism for equatorial ridge formation on spinning-top-shaped asteroids.
- We calculated the distribution of ejecta blankets on an ideal sphere with the parameters of Ryugu.
- Ejecta emplacement can feed equatorial ridges within the lifetime of an asteroid.

**Abstract**

The Japanese spacecraft Hayabusa 2 visited the asteroid (162173) Ryugu and provided many high-resolution images of its surface, revealing that Ryugu has a spinning-top shape with a prominent equatorial ridge, much like the shapes reported for some other asteroids. In this study, through dozens of numerical calculations, we demonstrate that during a period of fast rotation, ejecta from craters formed at lower and mid-latitudes can be deposited on the equatorial ridge. Assuming a rotation period of 3 h, we estimate that an equatorial ridge with a height of 50 m can be generated in $128^{+47}_{-27}$ My for a main-belt asteroid, or $3.1^{+4.2}_{-1.2}$ Gy for a near-Earth asteroid. Therefore, an equatorial ridge can form within the average mean collisional lifetime of a km-sized asteroid within the main belt, but not for near-Earth asteroids. Furthermore, our model may explain why blue (younger) material occurs on the equatorial ridge.

**1. Introduction**

Some asteroids have the shape of a spinning top, dominated by an equatorial ridge. The first clear case of a spinning-top-shaped asteroid was the primary of the binary asteroid (66391) Moshup (provisionally designated as 1999KW$_4$), whose three-dimensional shape was estimated using radar observations from the Arecibo Observatory and the Goldstone Solar System Radar (Ostro et al., 2006). Planetary radar astronomy and in-situ planetary explorations have found similar top-shaped asteroids, such as (311066) 2004DC (Taylor et al., 2008), (341843) 2008EV$_5$ (Busch et al., 2011), (136617) 1994CC (Brozovic et al., 2011), (308635) 2005YU$_{55}$ (Busch et al., 2012), the primary of (185851) 2000DP$_{107}$ (Naidu et al., 2015), (101955) Bennu (Nolan et al., 2013; Lauretta et al., 2019), and (162173) Ryugu (Watanabe et al., 2019). In particular, Ryugu and Bennu are the first asteroids to be visited by a spacecraft able to transmit large amounts of data to Earth. Interestingly, the Hayabusa 2 and OSIRIS-Rex spacecraft revealed that the equatorial ridges of both asteroids are heavily cratered, indicating that these equatorial ridges are fossil structures (Walsh et al., 2019; Sugita et al., 2019; Hirata et al., 2020).

Various formation mechanisms have been suggested for equatorial ridges. It is known that the so-called Yarkovsky– O'Keefe–Radzievskii–Paddack (YORP) effect changes the rotation state of sub-km-radius asteroids (spinning them up or down) through recoil from the emission of absorbed sunlight (Rubincam, 2000); for example, this effect has been detected as a spin-up of the rotation rate of (54509) YORP (Lowry et al., 2007; Taylor et al., 2007), (1862) Apollo (Kaasalainen et al., 2007), (1620)

Geographos (Durech et al., 2008), and Bennu (Hergenrother et al., 2019). Ostro et al. (2006) suggested that the primary of the 1999KW$_4$ system spins fast enough that an area of lowest potential is located at its equator; therefore, particles can freely move across its surface and would naturally seek out its equator as the lowest-energy state. Similarly, Scheeres et al. (2006) suggested that, because the minimum geopotential of the primary of the 1999KW$_4$ system is located along the equator, loose material preferentially migrates toward the equator owing to a type of mass-movement process (e.g. Scheeres et al., 2016, 2019). Walsh et al. (2008) and (2012) performed numerical simulations of YORP spin-up with a cohesionless body consisting of numerous rigid spheres and demonstrated that (i) a critically spinning spherical body becomes oblate through mass movement from its poles to its equator, (ii) mass shed from the equator accretes into a satellite (i.e., the secondary of the binary system), and (iii) subsequent to mass loss from its equator, an equatorial belt of material remains on the body. On the other hand, Hirabayashi and Scheeres (2014) demonstrated that (i) a critically spinning spherical body begins to fail structurally in its center, (ii) the material flow travels outward along the equatorial plane in the internal core, and (iii) this flow would result in an equatorial ridge around the body (e.g., Hirabayashi and Scheeres, 2019; Hirabayashi et al., 2015; Zhang et al., 2019; Watanabe et al., 2019). Michel et al. (2018), (2019), and (2020) proposed that an equatorial ridge could be formed through reaccumulation after a catastrophic disruption of a parent body; this scenario does not require a critically spinning body or YORP spin-up.

In this paper, we propose ejecta accumulation on a rapidly spinning object as the possible origin of an equatorial ridge. We calculate the global distribution of the ejecta thickness from an impact crater on a Ryugu-equivalent spherical object, with a radius of 448.2 m and a mass of $M = 4.5 \times 10^{11}$ kg (Watanabe et al., 2019), and demonstrate that ejecta preferentially accumulates along the equator.

## 2. Method

To compute the distribution of ejecta, we followed the method described by Hirata et al. (2021). We briefly describe the method in this section. The initial launch velocity ($v_{ej}$) of an ejecta particle is given by Housen and Holsapple (2011):

$$v_{ej} = C_1 \left( H_1 \sqrt[3]{\frac{4\pi}{3}} \right)^{-\frac{2+\mu}{2\mu}} \sqrt{gR_c} \left( \frac{x}{R_c} \right)^{-\frac{1}{\mu}} \left( 1 - \frac{x}{n_2 R_c} \right)^p \quad , (n_1 a \leq x \leq n_2 R_c) \quad (1)$$

where $v_{ej}$ is the initial launch velocity of a particle in the gravity regime, $R_c$ is the apparent crater radius, $x$ is the distance between the ejection position and the impact point, $g$ is the surface gravity, $a$ is the projectile radius, and the rest are scaling constants. Housen and Holsapple (2011) provided eight sets of scaling constants, while we selected a set applicable to dry sand ($\mu = 0.41, C_1 = 0.55, H_1 = 0.59, n_1 = 1.2, n_2 = 1.3, p = 0.3$). This is reasonable because Arakawa et al. (2020) suggested that the surface of Ryugu is composed of cohesionless sand-like material and its cratering occurs in a gravity-dominated regime, based on an artificial impact crater experiment. The initial launch position $x$ of the particle and the launch velocity $v_{ej}$ from the launch position is defined by the point at which the ejecta particle crosses the original surface. Distances such as $x$ and $R_c$ are defined as great-circle distances on a sphere. It should be noted that $R_c$ is the apparent crater radius, and $n_2 R_c$ is the crater rim radius (Figure 1). In general, the initial launch velocity and volume of the ejecta are axially symmetric and depend on the distance from the crater center. The ejecta launch angle was set to 45° for every ejecta particle (Housen and Holsapple, 2011), which is consistent with the angle of the impact ejecta curtain observed on Ryugu (Arakawa et al., 2020). We assumed that $n_1 a = 0$, because the size of the projectile was very small compared to the crater.

We computed the total volume $V$ of ejecta inside $x$, i.e., the ejecta with an initial velocity greater than $v_{ej}(x)$, using the equation:

$$V = kx^3, \quad (0 \leq x \leq n_2 R_c) \quad (2)$$

where $k$ is a constant. Housen and Holsapple (2011) suggested that $k = 0.3$ in the case of dry sand. To discretize Eq. (2), we assumed that the ejecta volume at launch was also axisymmetric. The volume of a particle launched at $(x_i, \theta_j)$ in polar coordinates is represented by the volume of ejecta launched from a small area bounded by a radial distance between $x_{i+1}$ and $\Delta \theta$:

$$V_{i,j} = k \frac{\Delta \theta}{2\pi} [x_i^3 - (x_i - \Delta x)^3] \quad (3)$$

where

$$x_i = i \Delta x, \quad \Delta x = \frac{n_2 R_c}{N_1}, \quad \Delta \theta = \frac{2\pi}{N_2}, \quad i = \{1, \ldots, N_1\}, \quad j = \{1, \ldots, N_2\} \quad (4)$$

In this study, we set $N_1 = 1000, N_2 = 360$.

Assuming that the object rotates with an angular velocity given by the vector $\mathbf{\Omega} = \{0,0,\omega\}$, the equation of motion is given by Scheeres et al. (1996):

$$\ddot{\mathbf{r}} + 2\mathbf{\Omega} \times \dot{\mathbf{r}} + \mathbf{\Omega} \times \mathbf{\Omega} \times \mathbf{r} = -\frac{GM}{|r|^3}\mathbf{r} \quad (5)$$

where $\mathbf{r}$ is the position vector of a particle relative to the asteroid center. The rotation rate is expressed as $\omega = 2\pi/(T \times 3600)$ rad/s, where T is in hours. The z-axis was set to the rotational axis, and the xy-plane was taken as the equatorial plane of the object.

Given the initial launch position and velocity of the particle, as described by Eq. (1), the trajectory of the particle was obtained by numerically integrating Eq. (5) until one of three outcomes occurred: a) the particle reached an altitude greater than 700 asteroid radii (i.e., $|r| > 168R_a$, where $R_a = 448.2$ m); b) the particle was below the asteroid surface; or c) the particle had revolved around the asteroid more than five times. We conducted calculations in 1-s steps. When the particle landed on the asteroid surface (i.e., $|r| < R_a$), the location (latitude and longitude) was recorded. To obtain the ejecta thickness, the sum of the volume of particles within a 1° colatitude circle was divided by the area of the 1° colatitude circle.

Note that we did not consider secondary ejection or lateral movements after landing, such as subsequent downslope motion. Owing to this assumption, the calculated ejecta thickness was slightly larger than the actual thickness. In addition, we note that our analysis does not take topographic effects into account; as an equatorial ridge grows, the asteroid shape changes and becomes less spherical, affecting the distribution of ejecta from each crater and the generation rate of the equatorial ridge. Because we cannot obtain the resultant shape of the asteroid in this model, our analysis cannot determine whether the resultant shape corresponds to the observed spinning-top shape of Ryugu.

## 3. Single crater case study

We examined the distribution of ejecta blankets from a single crater, as functions of the asteroid rotation period, crater location, and crater size.

We first considered a simple example: the distribution of ejecta from a crater formed at the equator with a diameter of 100 m, as a function of the rotation period. Figure 2A shows the ejecta distribution at a rotation period of T = 3, 3.5, 5, 7.627, and 10000 h. Here, T = 3 h is the critical limit of the rotation period, which was derived by equating the acceleration of gravity at the surface with the centrifugal acceleration at the equator of the 448.2 m spherical object with a Ryugu-equivalent mass. Roughly 80–90% of ejecta launched from a crater re-accumulated on the surface of Ryugu. At T = 10000 h, the ejecta thickness was symmetric, decreased gradually with increasing distance from the crater, and there was no sign of an equatorial ridge. In this case, the average height of the ejecta deposit at the equator was $3.9 \times 10^{-4}$ m. However, the ejecta distribution at a faster rotation period was different from that at T = 10000 h; the ejecta curtain was blown harder along the equator in the opposite direction of rotation. For example, at T = 3 h, the average height of the ejecta deposit at the equator was $1.2 \times 10^{-3}$ m, which is three times greater than that at T = 10000 h (Figure 2).

Furthermore, we examined the effect of latitudinal variation (crater location) on ejecta distribution. Figure 2B and 2C show the distributions of ejecta from craters formed at 30°N and 60°N, as a function of the rotation period. Figure 3 shows the total ejecta deposits in each 1° latitudinal band. In the case of craters formed at 0°N and 30°N (Figure 3A and 3B), the total volume of ejecta particles in the equatorial region increased as the rotation period increased; for example, there was a difference of two orders of magnitude between volumes at T = 3 h and 10000 h. In the case of craters formed at 60°N (Figure 3C), a slight increase in ejecta deposits occurred at the equator. Additionally, we found that the distribution of ejecta from craters located outside of the equatorial region tends to have two peaks: one around the craters, and another at the opposite latitude. For example, a crater formed at 30°N exhibited ejecta distribution peaks at 30°N and 30°S. As a result, at faster rotation periods, ejecta from the low-latitude crater preferentially accumulated at lower latitudes, whereas ejecta from the high-latitude crater did not.

We calculated and compared the distributions of ejecta as a function of crater size, using diameters of 10, 30, 50, 100, and 200 m. We set a rotation period of T = 3 h and the crater location to 0°N, 180°E. The profiles shown in Figure 4 were taken along the longitudinal line of 270°E. These profiles mostly show the same shape, and the only difference is that the amount of ejecta increases by a factor of approximately 10 as the crater diameter doubles. This tendency is consistent with the total volume of the ejecta shown in Eq. (2), which is proportional to the third power of the crater radius. Based on

Eq. (2), the total ejecta volume from 1000 craters with a diameter of 10 m should be almost equivalent to that of a single crater with a diameter of 200 m. According to Morota et al. (2020), the power law of the crater size frequency distribution is about −2.7 on NEAs (near-Earth asteroids) and MBAs (main-belt asteroids), which means that an average formation time of only one 200m-diameter crater on Ryugu almost corresponds to that of 1700 craters larger than 10m in diameter. Hence, smaller craters contribute more than larger craters to feeding the equatorial ridge.

## 4. Multiple craters case study

This section presents the ejecta distribution generated by multiple craters. Using a random-number generator, we generated numerous craters on a sphere with a radius of 448.2 m so that craters were uniformly distributed on the sphere and their sizes were determined by the crater production function. According to Morota et al. (2020), the power law of the cumulative size frequency distribution of the crater production function is −2.70 or −2.76, respectively, if we assume a collision-frequency model for an NEA or MBA. Figure 5 shows two examples, with values of −2.70 for the power law, 50 for the number of craters, 300 m and 50 m for the maximum and minimum crater diameters, respectively, and T = 3 h, 7.627 h, and 10000 h for the rotation periods of the object. Note that a diameter of 300 m roughly corresponds to the size of the largest crater on Ryugu, Urashima (290 m in diameter). Figure 6 shows the average profile of ejecta thickness between 145°E and 150°E, as well as the profiles at T = 3.1, 3.2, 3.3, 3.4, and 3.5 h (setting the same location, size, and number of craters as in Figure 5). In Figure 6A, ejecta thickness at the equator is ~1.5 m when T = 3 h, ~0.5 m when T = 3.2 h, and below 0.1 m when T = 10000 h. The ejecta did not concentrate at the equatorial region at T = 10000 h or 7.627 h, but concentration in the equatorial region did occur at approximately T < 3.2 h. Perhaps when the rotation period is close to the critical limit, the ejecta from equatorial craters, which usually accumulate near the crater rim under higher rotation periods, spread across the entire equator.

Furthermore, we calculated and compared 10 cases with varying parameters, as shown in Tables 1 and 2. In Cases A, B, C, D, and E, an NEA collision-frequency model was assumed, while Cases F, G, H, I, and J assumed an MBA collision-frequency model. In each case, we used 300 m as the maximum diameter and T = 3 h as the rotation period, while the minimum diameter of the craters was varied. We then set the number of craters so that the crater age corresponds to 3 My for cases A to E (NEA models) and to 100 My for cases F to J (MBA models). The crater ages were based on the collision-frequency model provided by Morota et al. (2020). In general, an MBA is more frequently cratered than an NEA. For simplicity, we defined the average thickness of the ejecta along the equator as the resultant height of the equatorial ridge, although it is affected by crater rims and depressions. We calculated 10 trials for each model using different sets of random numbers. We then obtained the average and standard deviation of the heights of the equatorial ridges from the 10 trials (Tables 1 and 2). For example, ejecta accumulation on an NEA generated a ridge at a rate of ~0.01 m per My for case A, while on an MBA, the ridge formed at a rate of 0.3 m per My for case F. However, these rates depend on the minimum diameter of craters; for example, the rate was ~0.01 m per My for case A (min. diameter = 10 m) but ~0.005 m per My for case E (min. diameter = 100 m). This indicates that the contribution of small craters is not negligible, as discussed in Section 3.

## 5. Discussion

The mean and equatorial radii of Ryugu are 448 m and 502 m, respectively (Watanabe et al., 2019), hence we assume that an ejecta thickness of 50 m is required to explain Ryugu's equatorial ridge. The generation rates in models A and F can feed enough ejecta deposits to the equator in $125^{+47}_{-27}$ My for an MBA and $3.1^{+4.2}_{-1.2}$ Gy for an NEA. The mean collisional lifetime for a km-sized asteroid is approximately 300–500 Ma (O'Brien and Greenberg, 2005; Bottke et al., 1994). Therefore, an equatorial ridge can be fed sufficiently within the average mean collisional lifetime of a km-sized asteroid within the main belt. It has been suggested that Ryugu used to be an inner MBA, such as the Eulalia or Polana asteroid family, possibly via more than one generation of parent bodies (Sugita et al., 2019). Bottke et al. (2015) estimated the breakup times of (142) Polana and (495) Eulalia as $1400^{+150}_{-150}$ Ma and $830^{+370}_{-150}$ Ma, respectively. In this scenario, the formation of Ryugu's equatorial ridge should have been mostly completed by the time Ryugu migrated from an inner MBA to NEA orbit.

The formation process of the equatorial ridge in our model is associated with the cratering process; therefore, it is consistent with the observed nature of the heavily cratered (i.e., geologically old) equatorial ridges on Ryugu and Bennu. In addition, our model explains the distribution of bluer materials on Ryugu. Despite the advanced age of the ridge, indicated by the existence of numerous craters, the ridge exhibits a bluish color, indicating that it is covered with fresh materials (Morota et al., 2020). The blue color of the equatorial ridge may imply ejecta accumulation from small, relatively fresh craters toward the equator (Hirata and Ikeya, 2021).

The timescales required to form the equatorial ridge in this study require a constant rotation period of T = 3.0 h. However, on a geological timescale, asteroids do not rotate at a constant rate; for instance, the rotation period of Bennu is currently accelerating at a rate of $3.63 \pm 0.52 \times 10^{-6}$ degrees day$^{-2}$, as a result of the YORP effect (Hergenrother et al., 2019). Although estimating the degree of influence of the YORP effect is very challenging, the YORP-cycle timescale, for asteroids with ~1 km diameters in the inner main belt, is assumed to be on the order of 10 to a few tens of My (e.g., Rubincum, 2000; Granvik et al., 2017). Therefore, if the equatorial ridge was formed by ejecta emplacement, it was likely fed gradually during the multiple periods during which Ryugu was a fast rotator.

It should be noted that these timescales ignore the contribution of craters smaller than 10 m, and therefore, we consider them as representing the minimum values. Based on Table 2, the accumulation rate in case F (min. diameter = 10 m) was twice as large as that in case J (min. diameter = 100 m). If we assume the minimum diameter to be 1 m, it is possible that the accumulation rate could become twice as large as our estimate in case F. Nonetheless, it is difficult to determine conclusively whether such small craters contribute to ejecta accumulation at the equator, because Ryugu is covered with many boulders which have a so-called armoring effect. Consequently, an ejecta blanket from a submeter-sized crater would be very different to that from a large crater. Furthermore, the calculated values were obtained using a constant rotation period of T = 3.0 h. Based on the view in Figure 6, the time required for ridge formation may increase threefold, given T = 3.2 h, and ridge formation will not take place if T > 3.2 hours. In addition, this model ignores the irregular shape of the asteroid. The critical rotation period is 3 h in the case of a sphere with a radius of 448.2 m, while it is 3.59 h in the case of an equatorial radius of 502 m. Therefore, it is possible that the suitable rotation period for efficient equatorial ridge generation may become longer as the equatorial ridge grows.

Lastly, as we stated in Section 2, our model cannot obtain the resultant shape of the asteroid, and therefore, we cannot determine whether the asteroid shape predicted from this model corresponds to the observed shape of Ryugu. This model is too simple to examine the roles of secondary ejection, lateral movements after landing, and topographic effects caused by ridge growth. In order to obtain the asteroid shape predicted by our ejecta emplacement hypothesis and assess the similarity between the predicted and observed shapes, further numerical simulations will be required, such as N-body simulation.

## 6. Conclusion

Our calculations reveal that ejecta emplacement plays a significant role in equatorial ridge formation. Ejecta from craters formed at low and mid-latitudes accumulate mostly on the equator if the rotation period is sufficiently fast. For example, on a critically spinning spherical body with a radius of 448.2 m, impact ejecta can build an equatorial ridge with a height of 50 m in 3.1 Gy if we assume an NEA collision-frequency model, and in 128 My if we assume an MBA collision-frequency model. Each timescale is within the average mean collisional lifetime for a km-sized MBA, but not for an NEA; therefore, Ryugu likely obtained its prominent equatorial ridge before it migrated from an MBA to NEA orbit. Note that these estimated timescales required to form the ridge ignore the contribution of craters smaller than 10 m. If we assume the minimum diameter to be 1 m, it is possible that the accumulation rate could become twice as large as our estimates. However, the role of small craters is difficult to determine conclusively because the ejecta blankets from submeter-sized craters would be very different to those from large craters. The formation process of the equatorial ridge in our model is associated with the cratering process; therefore, it is consistent with the observed nature of the heavily cratered (i.e., geologically old) equatorial ridges of Ryugu and Bennu. In addition, the ridge exhibits a bluish color, indicating that it is covered with fresh material, despite its age. Our model can explain the bluer color of the equatorial ridge through ejecta accumulation from relatively fresh, small craters toward the equator. Further numerical simulations, such as N-body simulation, would be suitable for future work aiming to assess the similarity between the predicted and observed asteroid shapes.

**Acknowledgments**

We wish to thank T. Morota for providing the calculation results of the collision frequency models for Ryugu and R. Suetsugu for helpful comments to numerical simulations. We appreciate an anonymous reviewer and their comments. We thank all members of the Hayabusa2 mission team for their support in data acquisition. This work was partly supported by JSPS Grants-in-Aid for Scientific Research Nos. 20K14538 and 20H04614 and the Hyogo Science and Technology Association (NH). The shape model of Ryugu is freely available via the Data ARchives and Transmission System (DARTS) at ISAS/JAXA.

**Figure Caption**

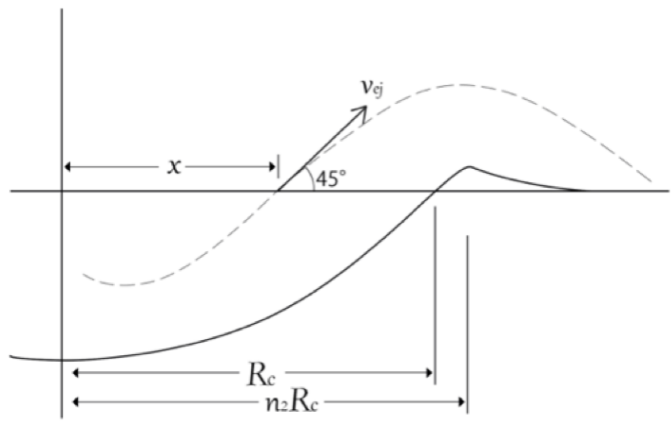

**Figure 1.** Definitions of parameters, adopted from Housen and Holsapple (2011), where $v_{ej}$ is the initial launch velocity of a particle, $R_c$ is the apparent crater radius, $n_2 R_c$ is the crater rim radius, and $x$ is the distance between the ejection position and the impact point.

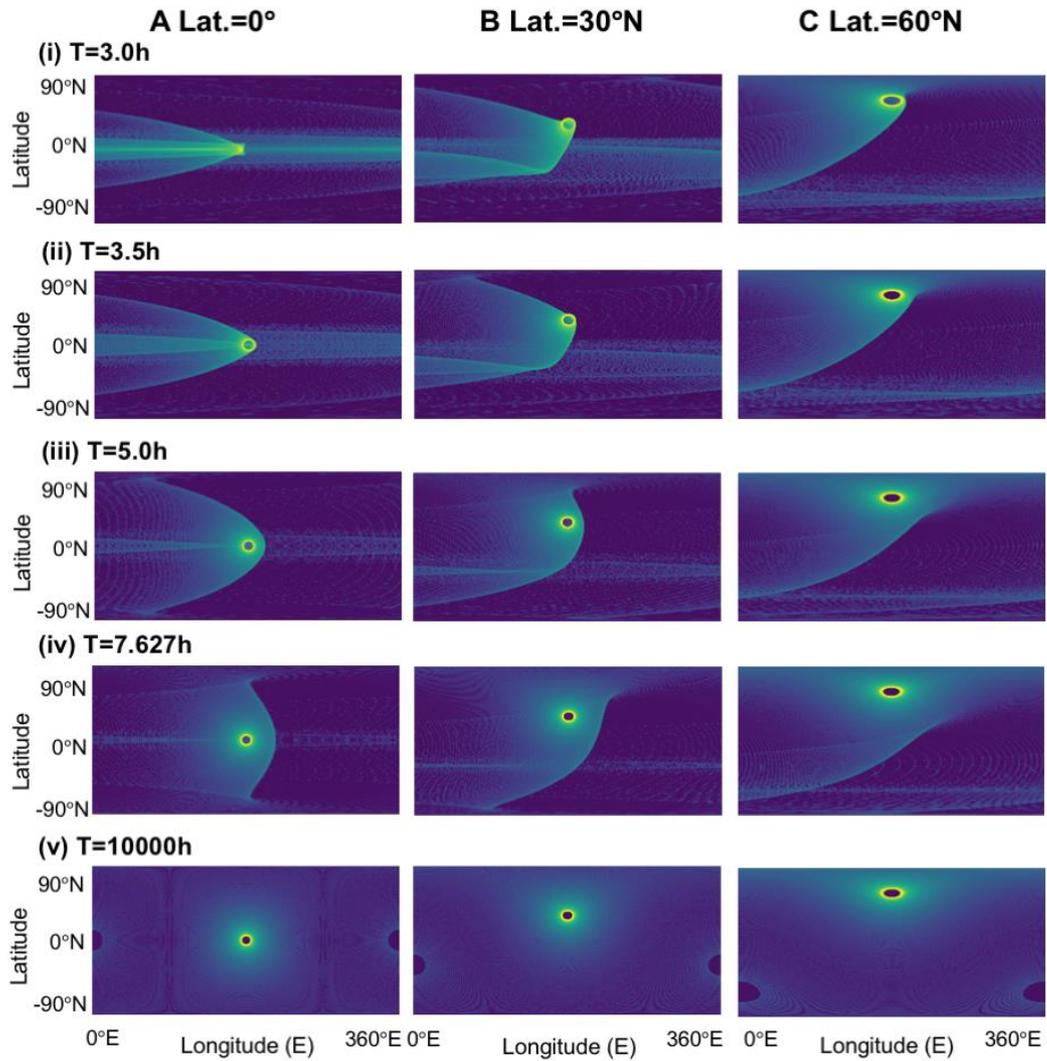

**Figure 2.** Global distribution of ejecta blankets as a function of crater and asteroid rotation period. Vertical and horizontal axes in each plate indicate latitude and east longitude, respectively. Ejecta thickness is colored using a log scale.

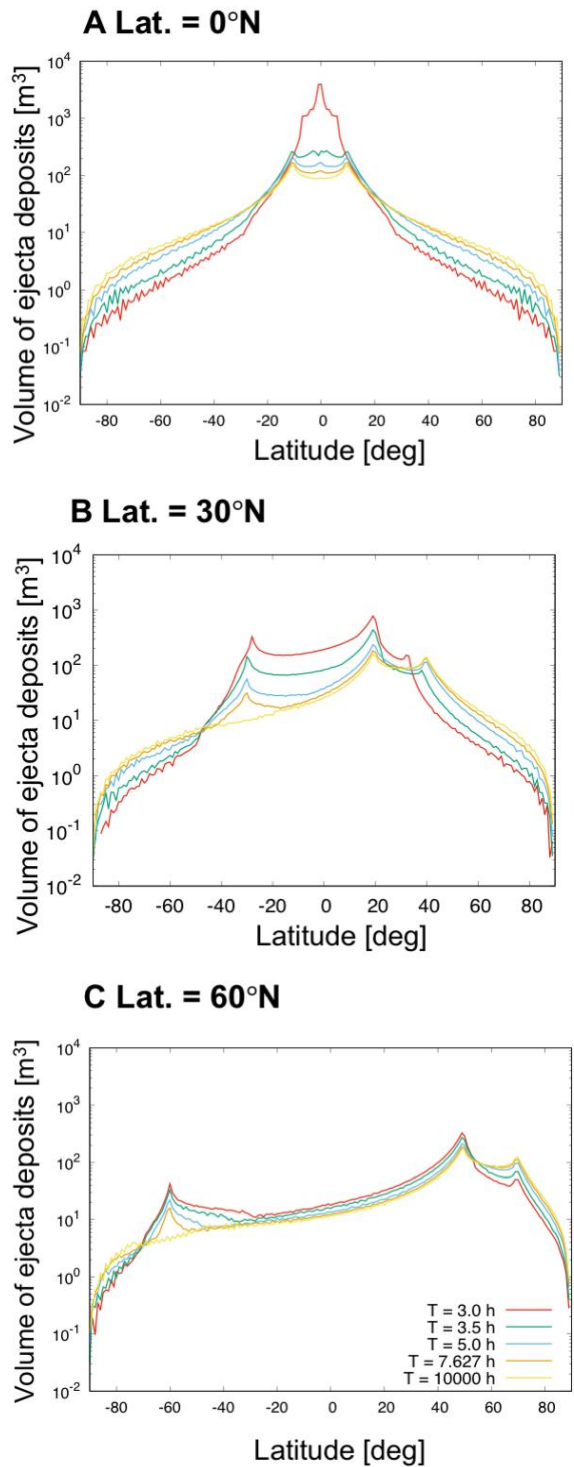

**Figure 3.** The latitudinal distribution of ejecta volume from craters formed at latitudes of (A) 0°N, (B) 30°N, and (C) 60°N as a function of the asteroid rotation period. The ejecta volume in each latitude indicates the total of ejecta volume falling within a 1-degree bin; i.e., a value at 30°N indicates the total volume of ejecta particles falling within a narrow latitudinal band between 29.5°N and 30.5°N. We excluded deposits that were within two crater radii from the crater center, to avoid including material in the crater rim.

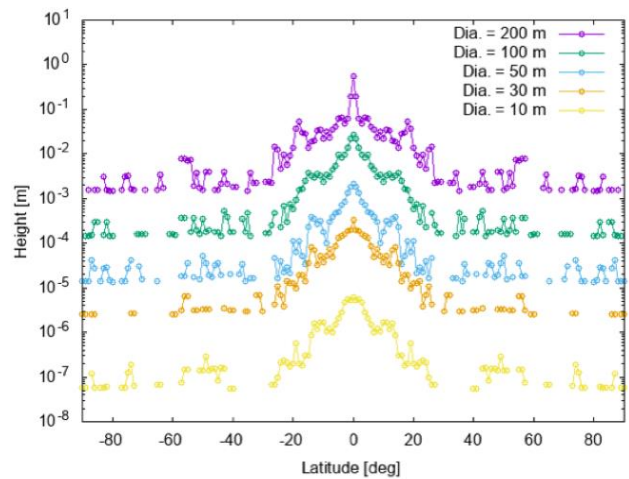

**Figure 4.** The profiles of ejecta thickness along the longitudinal line of 270°E as a function of crater diameter, ranging between 10 and 200 m, with a rotation period of 3 h and crater location of 0°N, 180°E. Because our calculations are discrete, the lines are disconnected.

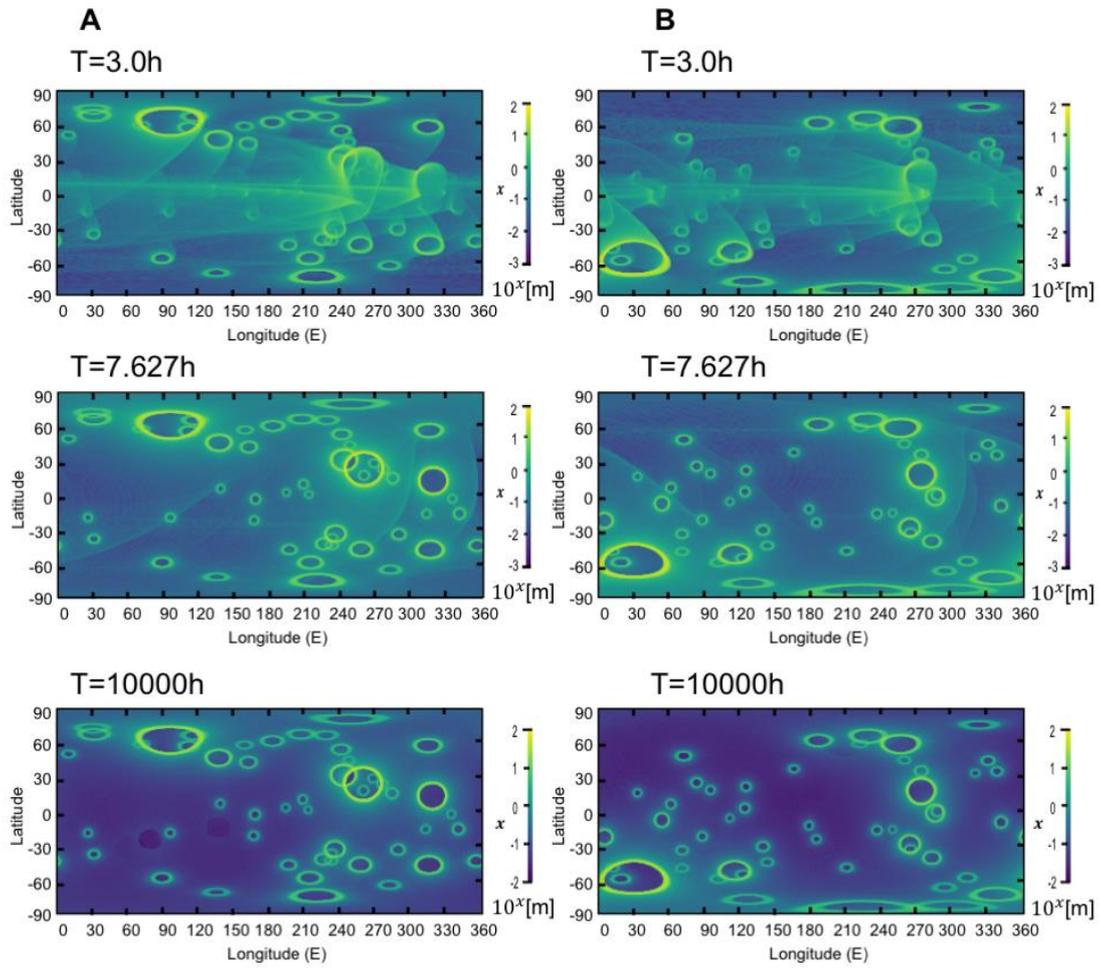

**Figure 5.** Global distribution of ejecta thickness generated by 50 randomly distributed craters, as a function of the rotation period (T = 3, 7,627, and 10000 h). We show two examples here (A and B columns), setting the same crater location and size but varying the rotation period.

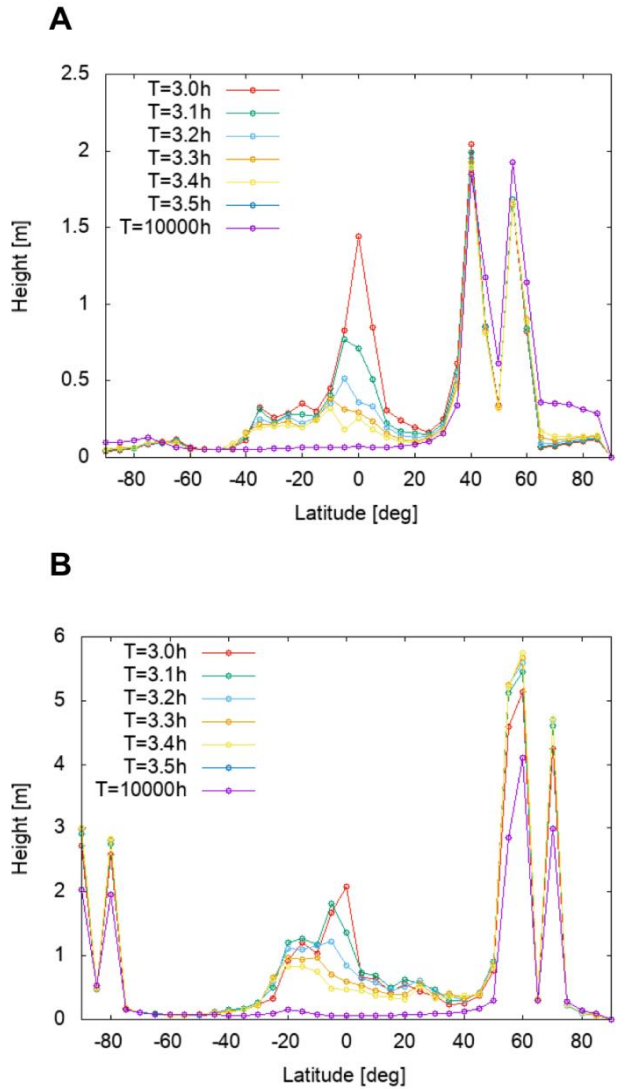

**Figure 6.** The profiles of ejecta thickness along different latitudes. The thickness indicated is an averaged value, between 150°E and 154°E (plate A) or 225°E and 229°E (plate B), in order to avoid large crater rims. The crater location and size in plates A and B corresponds to the A and B columns in Figure 5. Note that prominent peaks at 40°N and 60°N for A and 60°N, 70°N, and 80°S for B indicate crater rims.

**Table 1.** Conditions and resultant ridge heights of each NEA model[*].

| Case | A | B | C | D | E |
|---|---|---|---|---|---|
| **Max Dia. [m]** | 300 | 300 | 300 | 300 | 300 |
| **Min Dia. [m]** | 10 | 20 | 30 | 50 | 100 |
| **Number** | 3307 | 509 | 170 | 43 | 7 |
| **Averaged ridge height [m]** | 1.63 | 1.38 | 1.38 | 1.23 | 0.516 |
| **Standard deviation [m]** | 0.945 | 0.446 | 0.739 | 0.658 | 0.392 |
| **Ridge height generation rate [m/My]** | $(1.63 \pm 0.945) \times 10^{-2}$ | $(1.38 \pm 0.446) \times 10^{-2}$ | $(1.38 \pm 0.739) \times 10^{-2}$ | $(1.23 \pm 0.658) \times 10^{-2}$ | $(5.16 \pm 3.92) \times 10^{-3}$ |

[*] The collision frequency function model is for NEA, and the power law of crater size frequency distribution is -2.70. The number of craters corresponds to a crater age of 100 My on Ryugu.

**Table 2.** Conditions and resultant ridge heights of each MBA model[*].

| Case | F | G | H | I | J |
|---|---|---|---|---|---|
| **Max Dia. [m]** | 300 | 300 | 300 | 300 | 300 |
| **Min Dia. [m]** | 10 | 20 | 30 | 50 | 100 |
| **Number** | 3174 | 469 | 153 | 37 | 6 |
| **Averaged ridge height [m]** | 1.17 | 1.15 | 0.797 | 0.711 | 0.609 |
| **Standard deviation [m]** | 0.312 | 0.375 | 0.205 | 0.492 | 0.359 |
| **Ridge height generation rate [m/My]** | $0.390 \pm 0.104$ | $0.384 \pm 0.125$ | $0.266 \pm 0.068$ | $0.237 \pm 0.164$ | $0.203 \pm 0.120$ |

[*] The collision frequency function model is for MBA, and the power law is -2.76. The number of craters corresponds to a crater age of 3 My on Ryugu.